\documentclass[12pt]{article}
\usepackage{axodraw}
\title{{\normalsize
\begin{flushright}
hep-th/0601168
\end{flushright}} \vskip 2cm
\centerline
\bf Decay of  Spin-One Particle into Two Photons in 
Presence of Uniform External Magnetic Field}
\bigskip
\author{Dilip Angom, Kaushik Bhattacharya, Saurabh D. Rindani
\thanks{e-mail
addresses:angom@prl.res.in, kaushik@prl.res.in, saurabh@prl.res.in}\\
\normalsize
Theoretical Physics Division, 
Physical Research Laboratory, Ahmedabad 380009, India
}
\begin{document}
\def \ep1{{\bf e}_1}
\def \e2{{\bf e}_2}
\def \ev{{\bf e}_V}
\def \x{\times}
\def \b{{\bf b}}
\def \k{{\bf k}}
\def \d{\cdot}
\maketitle
\begin{abstract}
Yang's theorem states that an initial $J=1$ state cannot decay into
two photons. Because of this result some reactions relating to
elementary particles or atomic transitions can be ruled out. The
theorem is not valid in the presence of background electric or
magnetic fields.  In this work we show that the decay of a
$J=1$ particle into two photons is permitted by Bose symmetry and
rotational invariance when the background of the decay process is not
pure vacuum but contains an external classical magnetic/electric
field. We also discuss constraints on these amplitudes from {\bf CP}
invariance.
\end{abstract}
\section{Introduction}
It is well known that an initial $J=1$ state cannot decay into two
photons. This result is known in the literature by the name of Yang's
theorem or Yang-Landau theorem \cite{yang, Landau}. 
Yang's result is very general, and finds several
applications 
from atomic physics to elementary particle physics. In this article we
will only focus on decays of neutral elementary particles of spin
one. 
In the present article we investigate the decay of a neutral
elementary particle of spin one in the presence of a uniform classical
background magnetic field. Magnetic fields are easily produced in
laboratories and they are overwhelmingly present in the astrophysical
scenario, as in the core of a neutron star. If the initial decaying
particle has a magnetic moment it can interact with the magnetic
field.  But if the initial particle does not have any magnetic moment
magnetic fields can still enter into the picture through their
interaction with virtual charged fermions or other charged particles
propagating in the loops occurring in the Feynman diagram for the
decay process. There are many calculations employing suitable fermion
propagators, as for example the photon self-energy or photon pair
creation in a magnetic field \cite{Schwinger:1951nm, Chyi:1999fc,
Tsai:1974ap, Tsai:1974fa}.  The present article is completely general
in the sense that we have not specified any model of particle
interactions or carried out a loop integral to predict a unique result
for a specific situation. Our results are general and give some
important insight on the decay process in the presence of a magnetic
field.

In this article we have worked in the 3-vector language; the
calculations are not written in a Lorentz covariant fashion. The 
reason for this choice has been that a uniform classical magnetic
field implies a specific frame in which it is present. All the
calculations done are essentially specific to the frame where the
magnetic field exists and we have chosen that frame to be the rest
frame of the decaying boson. As the processes involve photons we
must be careful about the gauge invariance of the theory. In the
present case we work with photons in the Coulomb gauge where the
photon polarization vectors are transverse. 

Using Yang's theorem Gell-Mann predicted that the cross-section of
the reaction $\gamma \gamma \to \nu \overline{\nu}$ vanishes in the
four-Fermi limit \cite{Gell-Mann}. In the present article we show that
Yang's theorem does not hold true in the presence of an external magnetic
field and consequently the process $\gamma \gamma \to \nu
\overline{\nu}$ can happen. The calculation of the cross-section of
the above process to first order in the external magnetic field has
already been done by Shaisultanov \cite{Shaisultanov:1997bc}. This
results can have interesting astrophysical applications because the
reaction of two-photon decay to neutrinos is an efficient process of
energy dissipation from stars which does possess high magnetic fields.

The organization of the paper is as follows. Section \ref{yng} gives
the proof of Yang's theorem by enumerating amplitudes permitted by
rotational invariance and showing that they are ruled out by Bose
symmetry. In the subsequent section we specify the
most general terms which can constitute the decay amplitude in
presence of a magnetic field. By invoking symmetry arguments it will
be shown that only a few of the possible terms will actually contribute
for the two-photon decay in a magnetic field when {\bf CP} is a
symmetry of the theory. In the concluding section we will summarize
our results.
\section{Yang's theorem}
\label{yng}
%
\begin{figure}[h!]
\begin{center}
\begin{picture}(150,40)(0,-35)
\SetWidth{1}
\Photon(200,0)(90,0){2}{4}
\Text(207,0)[b]{$\gamma_2$}
\ArrowLine(180,-10)(200,-10)
\Text(190,-25)[b]{$\e2,\,\,-\k$}
\Vertex(85,0){5}
\Photon(-30,0)(80,0){2}{4}
\Text(-38,0)[b]{$\gamma_1$}
\Text(85,10)[b]{$V$}
\Text(85,-15)[b]{$\ev$}
\ArrowLine(-10,-10)(-30,-10)
\Text(-20,-25)[b]{$\ep1,\,\,\k$}
\end{picture}
\end{center}
\caption[]{The decay of a vector particle into two photons, $V \to \gamma_1\,
\gamma_2$. $\ev$, $\ep1$, $\e2$
are the polarizations of the initial vector particle and the two photons
respectively.}
\label{f:vgg}
\end{figure}
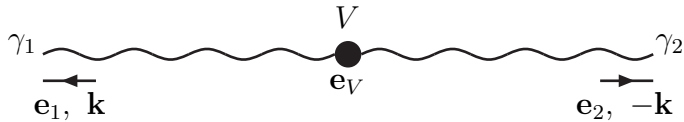
We give here a proof of Yang's theorem which is different from the
original proof given by Yang \cite{demille}.  Proof of the Yang's
theorem relies on the Bose symmetry of the two photons which are
produced after the decay. Fig.~\ref{f:vgg} shows the decay of a vector
particle $V$ into two photons $\gamma_1$ and $\gamma_2$. The initial
vector particle is assumed to be massive and we work in the rest frame
of $V$. In this case it must be possible to write the decay amplitude
${\cal M}$ for this process in terms of the quantities available: the
polarization of the two photons $\ep1$, $\e2$; the polarization of
$V$, $\ev$; and the outgoing photon momentum $\k=\k_1= -\k_2$. For the
subsequent proof of the Yang's theorem it does not matter whether we
designate $\ep1$ and $\e2$ as linear or circular polarization vectors
and so we will not specify the basis in which the polarizations of the
photons are supplied. Similarly the proof of Yang's theorem is
insensitive to the way the decaying particle is polarized, for all
choices and all basis of polarizations the following proof will hold
true.

On quantum field theoretical grounds we assume that the decay
amplitude ${\cal M}$ can contain the various polarization vectors only once but
any power of $\k$ can be present in the expression. In addition to
this, ${\cal M}$ cannot contain any factors of the form ${\bf
e}_{1,\,2}\cdot {\k}$ because of the transversality condition ${\bf
e}_{1,\,2}\cdot \k=0$.  In writing the expression for the
possible structure of ${\cal M}$ we can take into
account the various discrete symmetries like parity ({\bf P}),
charge conjugation ({\bf C}) and {\bf CP}. As the decay amplitude
consists of two factors of the final photon polarizations, as a
whole the amplitude will be insensitive to the way the photon field
changes due to the discrete symmetry transformations. The result will
only depend upon the way $\ev$ changes under the various discrete
transformations.

Let us suppose ${\ev}$ has well defined transformation properties
under the discrete transformations {\bf P} and {\bf C}.  If 
those transformations are symmetries of the theory, 
${\cal M}$
does not change after the application of
{\bf P}, {\bf C} on it. 
With this in mind we can have two separate
expressions for the amplitude, both of which respect {\bf P} and {\bf
C}. The reason for not writing one expression for the amplitude, which
could be a sum of the following two different forms of the amplitude,
is that the two possible forms of the amplitude as written
down do not have the same transformation properties under the discrete
transformations {\bf P}. It is assumed that the initial particle does
not have a definite {\bf C} transformation property or it transforms
trivially under {\bf C}. The two forms of the possible amplitude in
this case can be:
\begin{eqnarray}
{\cal M}(\ep1,\e2,\k) = (\ep1
\times \e2)\cdot \ev
\,F_1(k^2) + [(\ep1 \times \e2)\cdot
 \k](\ev
\cdot \k)\,F_2(k^2)\,,
\label{M1}
\end{eqnarray}
and,
\begin{eqnarray}
{\cal M}(\ep1,\e2,\k) = (\ep1 \cdot \e2)(\ev
\cdot \k)\,F_3(k^2)\,.
\label{M2}
\end{eqnarray}
$F_1(k^2)$, $F_2(k^2), ..$ are scalar functions which we will call the
form-factors. Eq.~(\ref{M1}) is the decay amplitude when the initial
particle transforms like a vector under parity while Eq.~(\ref{M2}) is
the decay amplitude when the initial particle transforms like a
pseudo-vector under parity. In both the above cases {\bf CP} is a
symmetry of the theory.

Now Bose symmetry implies that the decay amplitude will be the same if we
interchange the two final photons. This means,
\begin{eqnarray}
{\cal M}(\ep1,\e2,\k) =
{\cal M}(\e2,\ep1, -\k)\,.
\label{bs}
\end{eqnarray}
>From Eq.~(\ref{M1}) and Eq.~(\ref{M2}) it can be verified
that instead of being Bose symmetric
${\cal M}$ is antisymmetric under the two photon interchange.  As a result
the total decay amplitude which is:
\begin{eqnarray}
{\cal M} = {\cal M}(\ep1,\e2,\k) +
{\cal M}(\e2,\ep1, -\k) = 0\,.
\label{yangv}
\end{eqnarray}
As a result the decay $V \to \gamma_1\,\gamma_2$ is forbidden in
vacuum solely due to the fact that the terms available for making up
the amplitude ${\cal M}$ does not respect Bose symmetry.
\section{Two-photon decay of $J=1$ state in 
presence of a uniform background magnetic field}
\label{yngb}
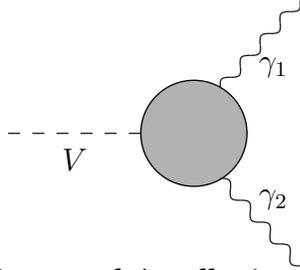
\begin{figure}[btp]
\begin{center}
\begin{picture}(180,120)(-90,-35)
\DashLine(-70,0)(-20,0){5}
\Text(-45,-10)[c]{$V$}
\GCirc(0,0){20}{.7}
\Text(30,25)[c]{$\gamma_1$}
\Text(30,-25)[c]{$\gamma_2$}
\Photon(10,17)(40,50)24
\Photon(10,-17)(40,-50)24
\end{picture}
\caption[]{\sf Feynman diagram of the effective vertex of $V$ with $\gamma_1$ 
and $\gamma_2$.
\label{f:effecv}}
\end{center}
\end{figure}
\begin{figure}[btp]
\begin{center}
\begin{picture}(180,120)(-90,-35)
\DashLine(-70,0)(-20,0){5}
\Text(-45,-10)[c]{$V$}
\BCirc(0,0){18}
\Text(-10,-27)[c]{$\ell$}
\Text(-10,28)[c]{$\ell$}
\Text(27,0)[c]{$\ell$}
\ArrowArc(0,0)(20,180,310)
\ArrowArc(0,0)(20,310,59)
\ArrowArc(0,0)(20,310,59)
\ArrowArc(0,0)(20,59,180)
\Text(30,25)[c]{$\gamma_1$}
\Text(30,-25)[c]{$\gamma_2$}
\Photon(10,17)(40,50)24
\Photon(10,-17)(40,-50)24
\end{picture}
\caption[]{\sf One-loop diagram contributing to the $V \to \gamma_1\,\gamma_2$
process where $V$ has tree-level interaction with charged leptons $\ell$. 
The double line propagators stands for the charged leptonic propagator in a 
magnetic/electric field.
\label{f:1la}}
\end{center}
\end{figure}
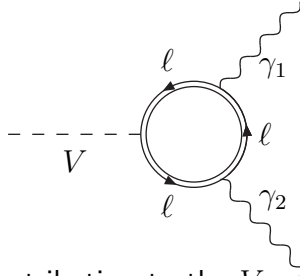

In the case where the background consists of a uniform magnetic field
vector ${\b}$, whose magnitude will be denoted by $b$, the analysis of
the previous section will change.  The most general Feynman diagram
for the process $V \to \gamma_1 \gamma_2$ is shown in
Fig.~\ref{f:effecv}. In our present work we do  not specify any
particular theory of particle interactions. Whatever we predict is
based on the most general assumptions of symmetry and transversality
of the photon fields. Consequently it is not possible for us to
specify the various interactions that can take place inside the shaded
blob of Fig.~\ref{f:effecv}. But if we are more specific and
allow $V$ to have interactions with the charged leptons in the standard
model, then, up to one-loop level, the decay of $V$ can proceed in the
way as shown in Fig.~\ref{f:1la}. The double line propagators
appearing inside the loop are the charged fermion propagators in the
presence of a background magnetic field first calculated by Schwinger
\cite{Schwinger:1951nm}\footnote{A general review on elementary
particle processes in presence of a constant background magnetic field
is present in \cite{Bhattacharya:2002aj}.}.

Adler has shown \cite{Adler:1971wn} that in a strong magnetic field
the dispersion relation of the photon changes and the waves 
do not strictly remain transverse. But in the present work we disregard
the effects of the modified photon dispersion relation and work
with transverse photons whose dispersion relations are unchanged from
that of the vacuum. To one-loop order our analysis is right, because changing
the dispersion relation of the photons essentially implies taking into
account 
two-loop effects where both the external photon legs get loop
corrections.

In the presence of a magnetic field the form of ${\cal M}$ will be guided
by the conditions of transversality and Bose symmetry, as it was in
vacuum.  In addition to these conditions we now have one more extra
condition that ${\cal M}$ should vanish when ${\b}\to 0$. The
particle $V$ is assumed to be massive and we work in its rest
frame, the frame where the magnetic field exists.

The various scalars constituted by the vectors at hand
can transform in different ways under the various discrete
transformations such as {\bf P}, {\bf C} and {\bf CP}. To classify those
scalars in terms of their transformation properties under the various
discrete transformations we have to assign some transformation
properties to ${\ev}$ i.e., specify how it transforms under
the various discrete transformations. Just for the sake of
completeness we also specify how the magnetic field vector $\b$
transforms under parity and charge-conjugation. Under parity,
\begin{eqnarray}
{\bf P}:\,\b \to \b\,,
\end{eqnarray}
that is, it transforms as a pseudo-vector. Under charge-conjugation,
\begin{eqnarray}
{\bf C}:\,\b \to -\b\,.
\end{eqnarray}
%
\subsection{When {\bf CP} is a symmetry of the theory}
\label{cpyngb}
For practical purposes we will present the case where $\ev$ does not
have any well-defined {\bf P} or {\bf C} transformation properties but
has a well defined {\bf CP} transformation property. This would be true, 
for example, for 
the $Z\to \gamma \gamma$ decay.

Suppose in the present case $\ev$ transforms under {\bf CP} as:
\begin{eqnarray}
{\bf CP}:\,{\ev} \to {\ev}\,.
\label{cpepv} 
\end{eqnarray}
With all the above restrictions and conditions taken into account the
various combinations of the vectors which give a scalar are written
down. The terms are grouped in two categories. The first division does
not contain any cross-products of vectors. The second division contains
all the terms which contain one cross-product among the various
available vectors.

Table 1 shows the various possible scalar terms made from the
available vectors at hand and the signs they acquire under the various
possible discrete transformations and the transformation involving the
interchange of two final photons, when ${\ev}$ transforms under {\bf
CP} as in Eq.~(\ref{cpepv}). The terms are arranged in such a manner
that all of them will have definite transformation properties when the
two final photons are interchanged because we expect the decay
amplitude to be Bose symmetric.
%
\begin{table}
\begin{center}
\begin{tabular}{|c|c|c|}
\hline
Terms &{\bf CP}& $\k \to -\k,\,\,\ep1 \leftrightarrow \e2 $\\
\hline
$(\ep1 \d \e2)(\ev \d \b)$& -  & +\\
$(\ep1 \d \e2)(\ev \d \k)$& - & -\\
$(\ep1 \d \b)(\e2 \d \b)(\ev \d \b)$& -  & + \\
$(\ep1 \d \b)(\e2 \d \b)(\ev \d \k)$& - & -\\
$(\ep1 \d \ev)(\e2 \d \b) + (\e2 \d \ev)(\ep1 \d \b)$& - & + \\
$(\ep1 \d \ev)(\e2 \d \b) - (\e2 \d \ev)(\ep1 \d \b)$& - & -\\
\hline
$(\ep1 \x \e2) \d \ev $& + & -\\
$[(\ep1 \x \e2)\d \k](\ev \d \b)$& + & +\\
$[(\ep1 \x \e2) \d \b](\ev \d \k)$& + & +\\
$[(\ep1 \x \e2) \d \k](\ev \d \k)$& + & -\\
$[(\ep1 \x \e2)\d \b](\ev \d \b)$& + & -\\
$[(\ev\x\k)\d\b](\ep1\d\e2)$& + & -\\
$[(\ev\x\k)\d\b](\ep1\d\b)(\e2\d\b)$& + & -\\
$[(\ep1 \x \ev)\d \k](\e2 \d \b)+ [(\e2 \x \ev)\d \k](\ep1 \d \b)$& + & -\\
$[(\ep1 \x \ev) \d \b](\e2 \d \b) + [(\e2 \x \ev) \d \b](\ep1 \d \b)$ & + & +\\
\hline
\end{tabular}
\caption{The possible constituents of the decay amplitudes, which are scalars
consisting of polarization vectors and the magnetic field. The 
transformation properties of these under ${\bf CP}$ and Bose symmetry are
indicated,  even and odd are represented as $+$ and $-$ respectively.}
\nonumber
\end{center}
\end{table}
%
In writing the above terms we have deliberately omitted scalar
products of two cross-products as $(\e2 \x \ev)\d (\ep1 \x \k)$ and
ordinary products as $[(\ep1 \x \e2) \d \b][(\ev \x \k) \d \b]$. By the
well known identity:
\begin{eqnarray}
\epsilon_{ijk}\,\epsilon_{lmn}&=&\delta_{il}\,(\delta_{jm}\,\delta_{kn}
- \delta_{jn}\,\delta_{km}) - \delta_{im}\,(\delta_{jl}\,\delta_{kn}
- \delta_{jn}\,\delta_{kl})\nonumber\\
&+&\delta_{in}\,(\delta_{jl}\,\delta_{km}- \delta_{jm}\,\delta_{kl})\,,
\label{decomp}
\end{eqnarray}
where $\epsilon_{ijk}$ is the 3-dimensional totally antisymmetric
Levi-Civita tensor. From the above equation one can see that all the
terms involving more than one cross-products can be reduced to sum of
terms containing scalar products of the individual vectors appearing
inside the original term containing more than one cross-product.
Moreover
we have omitted some terms such as $[(\ep1 \x \k) \d \b](\e2 \d
\ev)-[(\e2 \x \k) \d \b](\ep1 \d \ev)$, $[(\ep1 \x \ev)\d \k](\e2 \d
\b) - [(\e2 \x \ev)\d \k](\ep1 \d \b)$ etc. The reason is that they
can be written as linear combinations of other scalars listed in the
above table. A thorough analysis of the independence of the above
listed elements will be presented in the appendix.

Unlike the vacuum case, where all the form factors were functions of
$k^2$ alone, here the form factors can be functions of $k^2$, $b^2$
and $\k \d \b$. As a result the form factors are sensitive to the
discrete transformations in the present case. This fact can have an
interesting consequence. In the vacuum case if the scalars, made up by
the available vectors, were antisymmetric under Bose symmetry, we
could easily discard them. In the presence of a magnetic field if we come
across some scalars which are antisymmetric under Bose symmetry then
the form-factor multiplying it can also be antisymmetric under Bose
symmetry and as such the contribution from such a term cannot be
neglected. 

Ideally the amplitude of the process $V \to \gamma_1\,\gamma_2$ in
presence of a magnetic, when ${\ev}$ transforms under {\bf CP} as in
Eq.~(\ref{cpepv}), will be given by the sum of the fifteen terms
listed in the above table multiplied by their respective
form-factors. But we will see shortly that by symmetry considerations
we can eliminate many of the above listed terms and end up in a much
shorter list of possible terms which can contribute to the amplitude
of the decay process $V \to \gamma_1\,\gamma_2$.

As {\bf CP} is assumed to be symmetry of the theory, all the terms
in table 1 which change sign under {\bf CP} can be ruled out at
first. This is due to the fact that the form factors must be functions
of $\k\d\b$ which are {\bf CP} even and will not change sign under a
{\bf CP} transformation while the terms written in the table will
change sign and the amplitude will change sign, and {\bf CP} will not
remain a symmetry of the theory. So one remains with the remaining
nine possible candidates which can constitute the decay amplitude. If
the final amplitude has to be Bose symmetric then the terms in the
table which have the same signs under both {\bf CP} transformations and
the final photon interchange transformation will have their
corresponding form factors even in powers of $\k\d\b$. For the other
kind of terms the form factors will be odd functions of $\k\d\b$.  The
possible form of the amplitude will be:
\begin{eqnarray}
{\cal M}&=&{\cal M}(\ep1, \e2, \k) + {\cal M}(\e2, \ep1, -\k)\nonumber\\
&=&I^{(o)}_1(k^2,\k\d\b, b^2) (\ep1 \x \e2) \d \ev + I^{(e)}_2(k^2,\k\d\b, b^2)[(\ep1 \x \e2)\d \k](\ev \d \b)\nonumber\\ 
&+&I^{(e)}_3(k^2,\k\d\b, b^2)[(\ep1 \x \e2) \d \b](\ev \d \k)+
I^{(o)}_4(k^2,\k\d\b, b^2)[(\ep1 \x \e2) \d \k](\ev \d \k)\nonumber\\
&+&I^{(o)}_5(k^2,\k\d\b, b^2)[(\ep1 \x \e2)\d \b](\ev \d \b) +
I^{(o)}_6(k^2,\k\d\b, b^2)[(\ev\x\k)\d\b](\ep1\d\e2)\nonumber\\
&+&I^{(o)}_7(k^2,\k\d\b, b^2)[(\ev\x\k)\d\b](\ep1\d\b)(\e2\d\b)\nonumber\\
&+&I^{(o)}_8(k^2,\k\d\b, b^2)\left\{[(\ep1 \x \ev)\d \k](\e2 \d \b)+ [(\e2 \x \ev)\d \k](\ep1 \d \b)\right\}\nonumber\\
&+&I^{(e)}_9(k^2,\k\d\b, b^2)\left\{[(\ep1 \x \ev) \d \b](\e2 \d \b) + [(\e2 \x \ev) \d \b](\ep1 \d \b)\right\}\,.
\label{compamp} 
\end{eqnarray}
The superscript $(e)$ and $(o)$ on top of the form-factors indicate
that they are `even' or `odd' functions in $\k \d \b$, a direct
consequence of the {\bf CP} symmetry of the theory. The above form of
the amplitude gives the magnetic field dependence of the process and
also shows that the decay is dependent on the direction of the emitted
photons momenta or the initial particle  polarization with respect to
the magnetic field.

\subsection{Some properties of the decay amplitude}
\label{pda}

As the decay takes place in the presence of a magnetic field the amplitude
${\cal M}$ will not be isotropic. The most general amplitude, however,
is too complicated to analyze because of the presence of 9 independent
terms. We therefore look at special situations which lead to some
simplification in the structure of the amplitude. 

In certain physical situations, 
it would be useful to consider the weak magnetic 
field limit of the amplitude. By weak field we mean
$eb\ll m^2$, where $e$ is the magnitude of the
electronic charge and $m$ may be the mass scale of a typical looping
fermion appearing in a diagram like the one shown in
Fig.~\ref{f:effecv}. 

It is clear that the amplitude in Eq.~(\ref{compamp}) vanishes in the
limit of $\b = 0$. If we keep terms up to linear order in $\b$, only
the first four terms in Eq.~(\ref{compamp}) survive. In drawing this
conclusion, we make use of the fact that the form factors $I_1^{(o)}$
to $I_4^{(o)}$ which are odd functions of $\k\cdot\b$ would at least
be linear in $\b$. We keep only the linear term from them.  We
conclude that in the linear approximation in the magnetic field, there
would be four independent amplitudes for the decay.

It is interesting to note that the amplitude in this case vanishes when
$\ep1$ and $\e2$ are parallel to each other.
If we take the photons to be linearly polarized and suppose the
particle 1, with polarization vector $\ep1$, to be traveling along
the positive $z$-axis, then the polarization vectors can be written as:
\begin{eqnarray}
\begin{array}{ccccc}
{\ep1}^{(1)}&=&\e2^{(1)}&=&(1,0,0)\,\\
{\ep1}^{(2)}&=&-\e2^{(2)}&=&(0,1,0)\,.
\end{array}
\label{pol2}
\end{eqnarray}
In the above equations the superscripts stands for the polarization
components. Using the above equations with Eq.~(\ref{cpper1}) it can
be seen that crossed polarizations of the photons are favored, whereas
parallel polarizations are disallowed in
the decay process. The polarizations of the photons can also be
specified in the circular basis which is given by:
\begin{eqnarray}
\begin{array}{ccc}
{\ep1}_L&=&\frac{1}{\sqrt{2}}\left({\ep1}^{(1)} + i{\ep1}^{(2)}\right)\,,\\
{\ep1}_R&=&\frac{1}{\sqrt{2}}\left({\ep1}^{(1)} - i{\ep1}^{(2)}\right)\,.
\end{array}
\label{circpol}
\end{eqnarray}
Similarly the left and right handed circular states can be defined for
the photon traveling along the negative $z$-axis. For the present
case we have,
\begin{eqnarray}
\begin{array}{ccccc}
{\ep1}_L&=&{\e2}_R&=&\frac{1}{\sqrt{2}}(1,i,0)\,\\
{\ep1}_R&=&{\e2}_L&=&\frac{1}{\sqrt{2}}(1,-i,0)\,.
\end{array}
\label{circpol2}
\end{eqnarray}
>From the above equations and Eq.~(\ref{cpper1}) it is seen that the
final photons can only be both left circularly polarized or both right
circularly polarized in the present case. 

A simplification in the amplitude can also occur if we
assume a realistic situation where the photons are emitted
perpendicular to the magnetic field in which case $\k\d\b=0$,
$(\ep1\x\e2)\d\b=0$ and all the odd functions of $\k\d\b$ vanish. In
this case the expression of the decay amplitude will look like
\begin{eqnarray}
{\cal M}_{\perp}&=&I^{(e)}_2(k_\perp^2,b^2)[(\ep1 \x \e2)\d \k_\perp]
(\ev \d \b)\nonumber\\
&+&I^{(e)}_9(k_\perp^2,b^2)\,\left\{[(\ep1 \x \ev) \d \b](\e2 \d \b) + [(\e2 \x \ev) \d \b](\ep1 \d \b)\right\}\,.
\label{cpper}
\end{eqnarray}
In the above equation $\k_\perp$ indicates that in the present case
the 3-momenta of the photons are perpendicular to the external
magnetic field.

We can make some more simplifications if we further take a weak magnetic field
limit discussed earlier.
In this limit it is worth while to look for terms
in the above amplitude which are proportional to $b$ alone. Now
$I^{(e)}_1(k_\perp^2,b^2)$ is an even function of $b^2$. If we assume
that it can be written as a power-series in $b^2$ then the first term
of the expansion can be assumed to be independent of $b^2$. The other
term in the expression of the amplitude in Eq.~(\ref{cpper}) has
higher powers of $\b$. If we denote the first term of the
power-series expansion of $I^{(e)}_1(k_\perp^2,b^2)$ as $f(k_\perp^2)$
then we can write Eq.~(\ref{cpper}) for a weak magnetic field as:
\begin{eqnarray}
{\cal M}_{\perp}=f(k_\perp^2)\,[(\ep1 \x \e2) \d \k_\perp](\ev \d \b)\,.
\label{cpper1}
\end{eqnarray}
%

%
%

Since this is a special case of the weak-field limit considered
earlier, the same conclusions about the polarization dependence of
amplitude hold.  Thus crossed linear polarizations of the two photons
would have the maximum probability.  Similarly, in the circular
polarization basis, the two photons would both be either left or right
circularly polarized.

Similarly one can calculate the general structure of the decay
amplitude when the final photons are parallel to the external magnetic
field.  

The above discussions on the most general grounds shows that
in presence of a magnetic field there can be decay processes where an
initial $J=1$ state can decay into two photons. In particular when the
emitted photons are perpendicular to the direction of the magnetic
field the expression of the decay amplitude simplifies. In this
case the decay amplitude is odd in the external field
$\b$. Particularly when we are looking at small field effects the
decay amplitude contains only one term as given in Eq.~(\ref{cpper1}).
\section{Conclusion}
\label{concl}
In this article we analyzed the two photon decay of a particle which
has an initial spin state $J=1$ in presence of a uniform external
classical magnetic field.  In absence of the external magnetic field
such a decay cannot occur because we cannot write any decay amplitude
which respects Bose symmetry. In the presence of an external magnetic
field we can in general write fifteen candidates for the decay
amplitude. If we apply further symmetries like {\bf CP}, the
number of candidates reduces to nine. In the weak-field approximation,
the number further reduces to four.

The decay amplitude shows anisotropic features. If we focus either on
the case where the magnetic field is weak, or where the emitted
photons travel perpendicular to the direction of the magnetic field
the amplitude simplifies. Finally, when the photons travel
perpendicular to the magnetic field, which is also assumed to be weak,
we could show that the amplitude becomes completely specified up to
only one unknown function which will depend upon a loop integral.

A specific application of our formalism would be to the decay
$Z\to\gamma \gamma$ decay.  In the approximation of weak magnetic
field, the amplitude in that case is expected to be proportional to
$\frac{e^3 b }{m^2}$ where $m$ is the scale of the mass of the
particle appearing in the loop.  An explicit calculation of this
process to first order in the magnetic field using fermion loops has
been done by Tinsley \cite{zgg}.  It is interesting to note that
though we have seen that in general four form factors contribute to
first order in $\b$, in the specific calculation of ref. \cite{zgg}
only three independent tensors contribute.  Moreover, the amplitude
has an additional factor of $m^2/m_Z^2$, making the amplitude
independent of the mass of the fermion appearing in the loop.

The analysis presented in the article can be also applied for the
cases where {\bf P} and {\bf C} are symmetries of the theory. Our
analysis of the decay process is completely general and the same
approach can be applied when instead of a magnetic field we have an
electric field. In that case the {\bf CP} transformation properties of
the scalars making up the amplitude will change accordingly.  In fact
all the terms which we have pointed out as candidates for making up
the decay amplitude in presence of a magnetic field are also valid in
presence of an electric field.

In the end we stress the fact that starting from no specific theory we
have predicted the general form of the $V\to \gamma \gamma$ decay
where the final photons travel perpendicular to the direction of the
magnetic field, assuming {\bf CP} symmetry, up to an unknown function.
The angular dependence of our result can be experimentally tested at
accelerators.
\appendix\section*{\hfil Appendix \hfil}
\section{Reduction of possible candidates for the decay amplitude}
\label{app}
In section \ref{cpyngb} it was mentioned that the terms written in the
table are the most general scalars which one can construct out of the
available vectors $\ep1$, $\e2$, $\ev$, $\k$ and $\b$ when ${\bf CP}$
is a symmetry of the theory and $\ev$ transforms under a ${\bf CP}$
transformation as in Eq.~(\ref{cpepv}). Moreover in trying to
construct the scalars from the above mentioned vectors one has to take
into account the transversality of the final photons and each scalar
must contain only one factors of $\ep1$, $\e2$ and $\ev$ from quantum
field theoretical ground.  It was pointed out in section \ref{cpyngb}
that in table 1, containing the scalars, we have omitted some
particular forms of scalars which contain scalar-products of two
cross-products or ordinary products of two cross-products because they
can be written in terms of scalar-products as shown in
Eq.~(\ref{decomp}). If we write down all possible scalar terms,
constituting the decay amplitude, which abide the above restrictions
the the constituent terms will be as given in table 2.
%
\begin{table}
\begin{center}
\begin{tabular}{|c|}
\hline
Terms\\ 
\hline
$(\ep1 \d \e2)(\ev \d \b)$\\
$(\ep1 \d \e2)(\ev \d \k)$\\
$(\ep1 \d \b)(\e2 \d \b)(\ev \d \b)$ \\
$(\ep1 \d \b)(\e2 \d \b)(\ev \d \k)$\\
$(\ep1 \d \ev)(\e2 \d \b)+ (\e2 \d \ev)(\ep1 \d \b)$\\
$(\ep1 \d \ev)(\e2 \d \b)-(\e2 \d \ev)(\ep1 \d \b)$\\
\hline
$(\ep1 \x \e2) \d \ev $\\
$[(\ep1 \x \e2)\d \k](\ev \d \b)$\\
$[(\ep1 \x \e2)\d \k](\ev \d \k)$\\
$[(\ep1 \x \e2)\d \b](\ev \d \k)$\\
$[(\ep1 \x \e2) \d \b](\ev \d \b)$\\
$[(\ev \x \k) \d \b] (\ep1 \d \e2)$\\
$[(\ev \x \k) \d \b](\ep1 \d \b)(\e2 \d \b)$\\
$[(\ep1 \x \k)\d \b](\e2 \d \ev) + [(\e2 \x \k)\d \b](\ep1 \d \ev)$\\
$[(\ep1 \x \k)\d \b](\e2 \d \ev) - [(\e2 \x \k)\d \b](\ep1 \d \ev)$\\
$[(\ep1 \x \ev) \d \k](\e2 \d \b)-[(\e2 \x \ev) \d \k](\ep1 \d \b)$\\
$[(\ep1 \x \ev) \d \k](\e2 \d \b)+[(\e2 \x \ev) \d \k](\ep1 \d \b)$\\
$[(\ep1 \x \ev) \d \b](\e2 \d \b)+[(\e2 \x \ev) \d \b](\ep1 \d \b)$\\
$[(\ep1 \x \ev) \d \b](\e2 \d \b)-[(\e2 \x \ev) \d \b](\ep1 \d \b)$\\
$[(\ep1 \x \k)\d \b](\ev \d \b)(\e2 \d \b) + [(\e2 \x \k)\d \b](\ev \d \b)(\ep1 \d \b)$\\
$[(\ep1 \x \k)\d \b](\ev \d \b)(\e2 \d \b) - [(\e2 \x \k)\d \b](\ev \d \b)(\ep1 \d \b)$\\
$[(\ep1 \x \k)\d \b](\ev \d \k)(\e2 \d \b) + [(\e2 \x \k)\d \b](\ev \d \k)(\ep1 \d \b)$\\
$[(\ep1 \x \k)\d \b](\ev \d \k)(\e2 \d \b) - [(\e2 \x \k)\d \b](\ev \d \k)(\ep1 \d \b)$\\
\hline
\end{tabular}
\caption{ The possible scalars consisting of the  polarization 
vectors and the magnetic field vector. Some of the terms are not
distinct as there are terms which are linearly dependent and 
this follows from Schouten identity.}
\end{center}
\end{table}
%

In table 2 all the scalars are not independent of one other, the 
reason being the Schouten identity for three dimensions, viz.,
\begin{eqnarray}
\epsilon^{ijk}\,p^l - \epsilon^{ljk}\,p^i - \epsilon^{ilk}\,p^j 
- \epsilon^{ijl}\,p^k=0\,,
\label{shuten}
\end{eqnarray}
where $p^i$ is an arbitrary 3-vector. With the help of the above
identity one can derive some related identities. If ${\bf A}$, ${\bf
B}$, ${\bf C}$, ${\bf D}$, ${\bf E}$, ${\bf F}$ and ${\bf G}$ are arbitrary
3-vectors then by the application of the above identity one can
derive,
\begin{eqnarray}
[({\bf A}\x{\bf B})\d{\bf C}]({\bf D}\d{\bf E})&=&[({\bf A}\x{\bf B})\d{\bf E}]({\bf C}\d{\bf D}) + [({\bf E}\x{\bf B})\d{\bf C}]({\bf A}\d{\bf D})\nonumber\\
&+&[({\bf A}\x{\bf E})\d{\bf C}]({\bf D}\d{\bf B})\,,
\label{id1}\\
\left[({\bf A}\x{\bf B})\d{\bf C}\right]({\bf D}\d{\bf E})({\bf F}\d{\bf G})
&=&
\left\{[({\bf A}\x{\bf B})\d{\bf E}]({\bf C}\d{\bf D}) +
[({\bf E}\x{\bf B})\d{\bf C}]({\bf A}\d{\bf D})\right.\nonumber\\
&+&\left.[({\bf A}\x{\bf E})\d{\bf C}]({\bf D}\d{\bf B})\right\}({\bf F}\d{\bf G})\,.
\label{id2}
\end{eqnarray}
By the application of Eq.~(\ref{id1}) one can show,
\begin{eqnarray}
[(\ep1 \x \k)\d \b](\e2 \d \ev) - [(\e2 \x \k)\d \b](\ep1 \d \ev)&=&
-[(\ep1 \x \e2)\d \k](\ev \d \b)\nonumber\\
&+&[(\ep1 \x \e2)\d \b](\ev \d \k)\,,
\end{eqnarray}
which implies the fifteenth term of table 2 is a linear sum of the
eighth term and the tenth term of the table. It is not an independent
scalar.

The sixteenth term of the last table can be written as:
\begin{eqnarray}
[(\ep1 \x \ev) \d \k](\e2 \d \b)-[(\e2 \x \ev) \d \k](\ep1 \d \b)&=&
-[(\ep1 \x \e2) \d \ev](\k \d \b)\nonumber\\
&+&[(\ep1 \x \e2) \d \k](\ev \d \b)\,,
\end{eqnarray}
which shows it is a linear superposition of the seventh and the eighth
terms of the table.  Similarly the nineteenth term of table 2,
\begin{eqnarray}
[(\ep1 \x \ev) \d \b](\e2 \d \b)-[(\e2 \x \ev) \d \b](\ep1 \d \b)
&=&-[(\ep1 \x \e2) \d \ev]\,b^2\nonumber\\ 
&+&[(\ep1 \x \e2) \d \b](\ev \d \b)\,, 
\end{eqnarray}
is a linear superposition of the seventh term and the twelfth 
term of the same table.

By the application of Eq.~(\ref{id2}) the twenty-first term can be 
decomposed as:
\begin{eqnarray}
& &[(\ep1 \x \k)\d \b](\ev \d \b)(\e2 \d \b) - [(\e2 \x \k)\d \b](\ev \d \b)(\ep1 \d \b)=\nonumber\\
& &~~~~~~~~~~~~~~~~~~~~~~~~~-[(\ep1 \x \e2)\d \k]\,b^2(\ev \d \b)
+ [(\ep1 \x \e2)\d \b](\k\d\b)(\ev \d \b)\,,
\end{eqnarray}
showing that it is a linear superposition of the eighth term and the
eleventh term of the table. Similarly the twenty-third term of table 2 
can be written as:
\begin{eqnarray}
& &[(\ep1 \x \k)\d \b](\ev \d \k)(\e2 \d \b) - [(\e2 \x \k)\d \b](\ev \d \k)(\ep1 \d \b)=\nonumber\\
& &~~~~~~~~~~~~~~~~~~~~~~~~~-[(\ep1 \x \e2)\d \k]\,b^2(\ev \d \k)
+[(\ep1 \x \e2)\d \b](\k\d\b)(\ev \d \k)\,,
\end{eqnarray}
which shows it is a linear superposition of the ninth and tenth terms
of the last table.

By the use of Eq.~(\ref{id2}) one can write,
\begin{eqnarray}
[(\ep1\x\k)\d\b](\ev\d\k)(\e2\d\b)=[(\ep1\x\k)\d\ev](\b\d\k)(\e2\d\b)
+ [(\ep1\x\ev)\d\b]k^2(\e2\d\b)\,,
\end{eqnarray}
and
\begin{eqnarray}
[(\e2\x\k)\d\b](\ev\d\k)(\ep1\d\b)=[(\ep1\x\k)\d\ev](\b\d\k)(\ep1\d\b)
+ [(\e2\x\ev)\d\b]k^2(\ep1\d\b)\,,
\end{eqnarray}
which implies the twenty-second term,
\begin{eqnarray}
& &[(\ep1\x\k)\d\b](\ev\d\k)(\e2\d\b)+[(\e2\x\k)\d\b](\ev\d\k)(\ep1\d\b)
=\nonumber\\ 
&&~~~~~~~~~~~~~~~~~~~~~-(\b\d\k)\left\{[(\ep1 \x \ev) \d \k](\e2 \d \b)+
[(\e2 \x \ev) \d \k](\ep1 \d \b)\right\}\nonumber\\
&&~~~~~~~~~~~~~~~~~~~~~+k^2\left\{[(\ep1 \x \ev) \d \b](\e2 \d \b)+[(\e2 \x \ev) \d \b]
(\ep1 \d \b)\right\}\,,
\end{eqnarray}
is a linear superposition of the seventeenth and eighteenth terms of table 2.
In the above derivation the transversality of the photon fields has been used.

Similarly it can be shown that term twenty,
\begin{eqnarray}
& &[(\ep1 \x \k)\d \b](\ev \d \b)(\e2 \d \b) + [(\e2 \x \k)\d \b](\ev \d \b)
(\ep1 \d \b)=\nonumber\\
& &2[(\ev \x \k) \d \b](\ep1 \d \b)(\e2 \d \b) - b^2\left\{[(\ep1 \x \ev) \d \k](\e2 \d \b)+[(\e2 \x \ev) \d \k](\ep1 \d \b)\right\}\nonumber\\
&+&(\b\d\k)\left\{[(\ep1 \x \ev) \d \b](\e2 \d \b)+[(\e2 \x \ev) \d \b](\ep1 \d \b)\right\}
\end{eqnarray}
is a linear superposition of term thirteen, seventeen and eighteen.

By the application of Eq.~(\ref{id1}) one can show that the fourteenth term,
\begin{eqnarray}
[(\ep1 \x \k)\d \b](\e2 \d \ev) &+& [(\e2 \x \k)\d \b](\ep1 \d \ev)=
2[(\ev \x \k) \d \b] (\ep1 \d \e2)\nonumber\\
&-&\left\{[(\ep1 \x \ev) \d \k](\e2 \d \b)+[(\e2 \x \ev) \d \k](\ep1 \d \b)\right\}\,,
\end{eqnarray}
is a superposition of the twelfth term and the seventeenth term.

From the above discussion one can omit fourteenth, fifteenth, sixteenth,
nineteenth, twentieth, twenty-first, twenty-second and twenty-third terms
of the last table as because they can be written in terms of the other
scalars appearing in the table. Omitting these terms will lead to the
final form of the table as presented in section \ref{cpyngb} of the
article.


\begin{thebibliography}{999}

\bibitem{yang} C.~N.~Yang,\,Phys.\,Rev.\,{\bf 77}, 242 (1950)

\bibitem{Landau}
L.~D.~Landau, 
Sov.\ Phys.\ Doklady {\bf 60}, 207 (1948).

\bibitem{Schwinger:1951nm}
J.~S.~Schwinger,
Phys.\ Rev.\  {\bf 82}, 664 (1951).

\bibitem{Chyi:1999fc}
T.~K.~Chyi, C.~W.~Hwang, W.~F.~Kao, G.~L.~Lin, K.~W.~Ng and J.~J.~Tseng,
Phys.\ Rev.\ D {\bf 62}, 105014 (2000)
[arXiv:hep-th/9912134].

\bibitem{Tsai:1974ap}
W.~y.~Tsai,
Phys.\ Rev.\ D {\bf 10}, 2699 (1974).

\bibitem{Tsai:1974fa}
W.~y.~Tsai and T.~Erber,
Phys.\ Rev.\ D {\bf 10}, 492 (1974).

\bibitem{Gell-Mann}
M.~Gell-Mann, 
Phys.\ Rev.\ Lett. {\bf 6}, 70 (1961).

\bibitem{demille}
D.~DeMille, D.~Budker, N.~Derr, and E.~Deveney,
in {\it Proceedings of the International Conference on Spin-Statistics 
Connection and Commutation Relations}
Edited by R. C. Hilborn and G. M. Tino, 
AIP Conf. Procs. No. 545, p. 227 (2000).
\bibitem{Shaisultanov:1997bc}
R.~Shaisultanov,
Phys.\ Rev.\ Lett.\  {\bf 80}, 1586 (1998)
[arXiv:hep-ph/9709420].

\bibitem{Bhattacharya:2002aj}
K.~Bhattacharya and P.~B.~Pal,
Proc.\,Ind.\,Natl.\,Sci.\,Acad.\, {\bf 70}: 145 (2004) 
arXiv:hep-ph/0212118.

\bibitem{Adler:1971wn}
S.~L.~Adler,
Annals Phys.\  {\bf 67}, 599 (1971).

\bibitem{zgg}
T.~M.~Tinsley,
Phys.\ Rev.\ D {\bf 65}, 013008 (2002)
[arXiv:hep-ph/0106142].

\end{thebibliography}
\end{document}